\documentclass{PoS}

\title{Hadronic models of high-energy radiation from microquasars: recent developments}

\ShortTitle{High-energy emission from microqusars}

\author{\speaker{Gustavo E. Romero}\thanks{Member of CONICET.}\\
        Instituto Argentino de Radioastronom{\'\i}a (CCT La Plata, CONICET),
C.C.5, 1894 Villa Elisa,  Buenos Aires, Argentina \\ and \\
Facultad de Ciencias Astron\'omicas y Geof{\'\i}sicas, Universidad Nacional de La Plata, Paseo del Bosque, B1900FWA La Plata, Argentina
\\
        E-mail: \email{romero@fcaglp.unlp.edu.ar}}

\abstract{If the jets of microquasars carry a significant power in the form of relativistic hadrons, then gamma rays and neutrinos can be produced by interactions with matter and photon fields either external or internal to the jet. In this paper I present some recent results of calculations of the interaction of hadronic jets with 1) matter of the jet itself, 2) photon fields generated by synchrotron radiation of both protons and electrons, and 3) matter external to the jet (e.g. a clumped wind). I briefly discuss neutrino production in these scenarios and the prospects of detection with new gamma-ray instruments. Finally, I make a few comments on the controversy about the nature of LS I +61 303.}

\FullConference{VII Microquasar Workshop: Microquasars and Beyond\\
		 September 1-5 2008\\
		 Foca, Izmir, Turkey}

\begin{document}

\section{Introduction}

Microquasars (MQs) are accreting binaries formed by a compact object and a donor star \cite{Mirabel1999}. The compact object can be a neutron star (e.g. as in Sco X-1) or a black hole (e.g. Cygnus X-1). The donor star can be of an early type or a low-mass star. In the first case we talk of High-Mass Microquasars (HMMQs), whereas in the second case of Low-Mass Microquasars (LMMQs).

MQs present non-thermal jets. This means that there are relativistic particles in the jets. It is not necessary for the jet itself to have a macroscopic relativistic outflow velocity, but in several cases bulk velocities with Lorentz factors $\Gamma\sim1.5 - 10$ have been inferred.   

In the environment of a MQ the presence of relativistic particles can result in the production of detectable gamma rays \cite{Levinson1996}.

Both leptonic and hadronic models have been proposed in the literature to predict the gamma-ray emission of MQs. A broad review which describes the bulk of the literature can be found in Ref. \cite{Ramon2008}. The reader is referred there for further references. 

A model, in the sense used here, is a conceptual mechanism that represents, as faithfully as possible, some of the physical processes occurring in the material object or system under study. In particular, a hadronic model for MQs is a model that represents radiative processes triggered by protons or other nuclei. There is not such a thing as a purely hadronic radiative model in astrophysics. All models are actually lepto-hadronic, since relativistic hadronic interactions unavoidably lead to meson production and the subsequent injection of leptons in the system. In the  following sections we will discuss the latest results on this type of models. We shall separate them according to the dominant hadronic process taking palace in the source. 

\section{Heavy jet models}

\begin{figure}
\includegraphics[width=.95\textwidth]{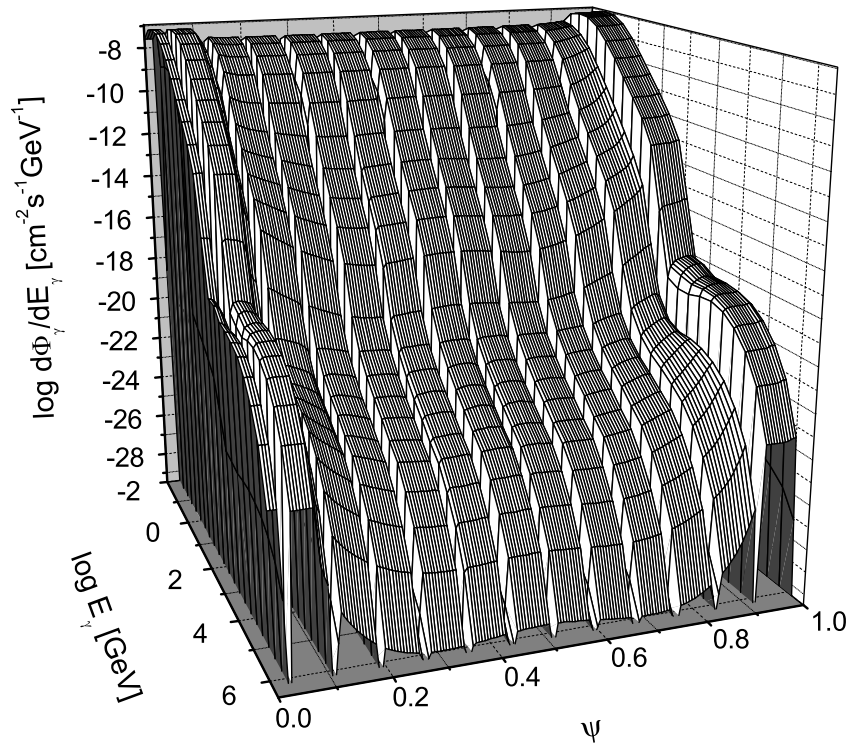}
\caption{Gamma-ray flux predicted by a heavy jet model for SS 433, as a function of time (orbital phase) and energy. Note the modulation produced by the variable absorption due to the precession of the jets \cite{Reynoso2008}.}
\label{fig1}
\end{figure}

\begin{figure}
\includegraphics[width=.95\textwidth]{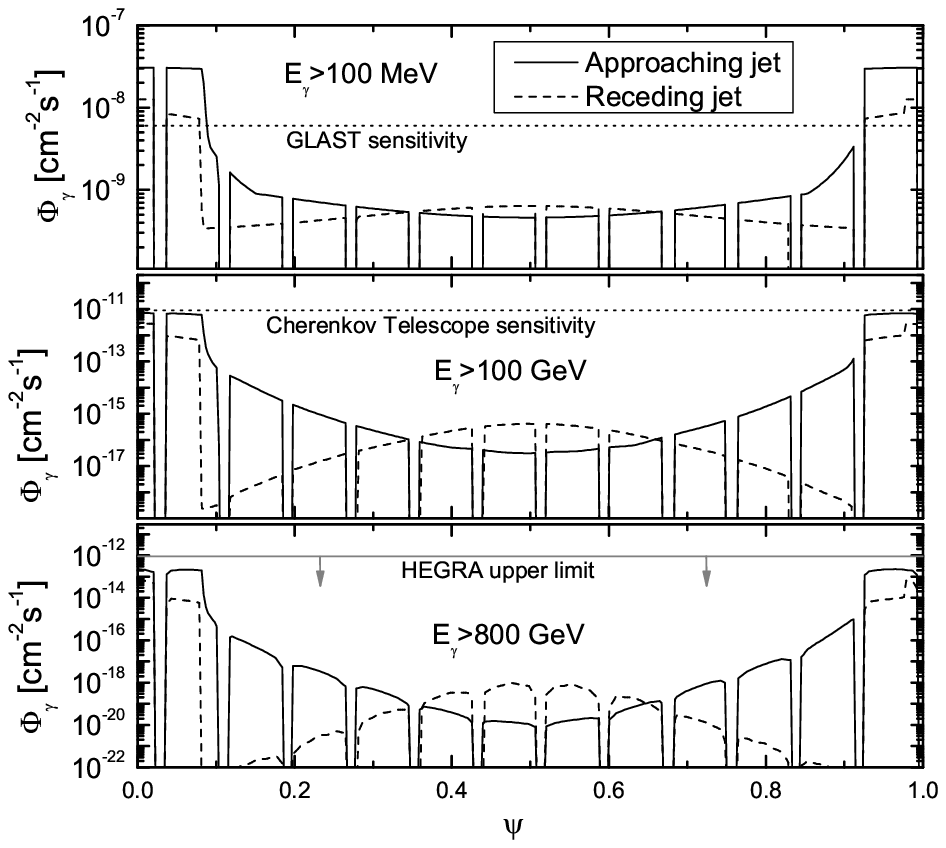}
\caption{Gamma-ray light curve from a heavy jet model of SS 433. The different panels show a comparison with the sensitivity of different instruments. The Cherenkov sensitivity of the middle panel corresponds to MAGIC \cite{Reynoso2008}.}
\label{fig2}
\end{figure}

These models assume that the jets of MQs have a very large kinetic energy and are dominated by cold barions. The original model was developed by Reynoso et al. \cite{Reynoso2008} and applied to SS 433, a source with extremely powerful jets (kinetic luminosity of $\sim 10^{40}$ erg s$^{-1}$ \cite{Dubner}). Gamma rays are mainly due to inelastic $pp$ collisions at the base of the jet, where the density of the thermal plasma is very high. Absorption effects in the ambient photon field produce a significant modulation of the emission \cite{Reynoso2008b} (see Figs. \ref{fig1} and \ref{fig2}).  

Other sources, like Cygnus X-1, seem to have heavy jets as well \cite{Heinz2006}, and a model where the high-energy emission is mainly due to  internal hadronic interactions might be appropriate. A crucial point is the fraction of relativistic protons in the outflow. GLAST observations can help to determine this point.

\section{The photo-hadronic jet model}

\begin{figure}
\includegraphics[width=.99\textwidth]{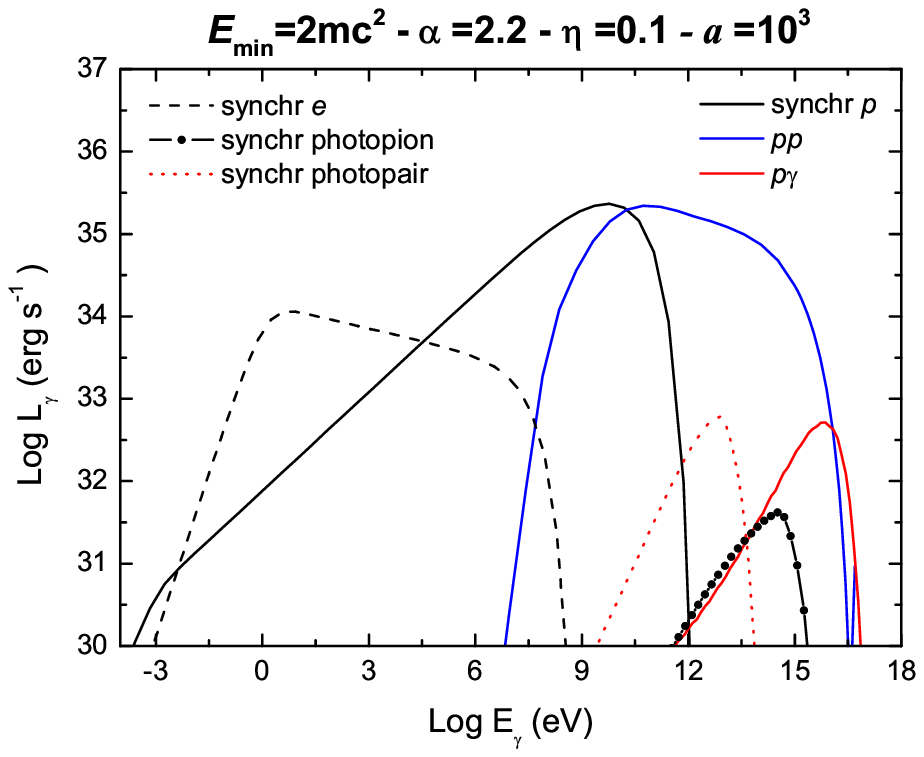}
\caption{Spectral energy distribution of a `proton' microquasar. In addition to the different components due to synchrotron and photo-hadron interactions, the internal $pp$ contribution is also shown \cite{Romero2009}. The assumed particle acceleration efficiency is 10\% and the injection is a power-law with index 2.2.}
\label{fig3}
\end{figure}

\begin{figure}
\includegraphics[width=.99\textwidth]{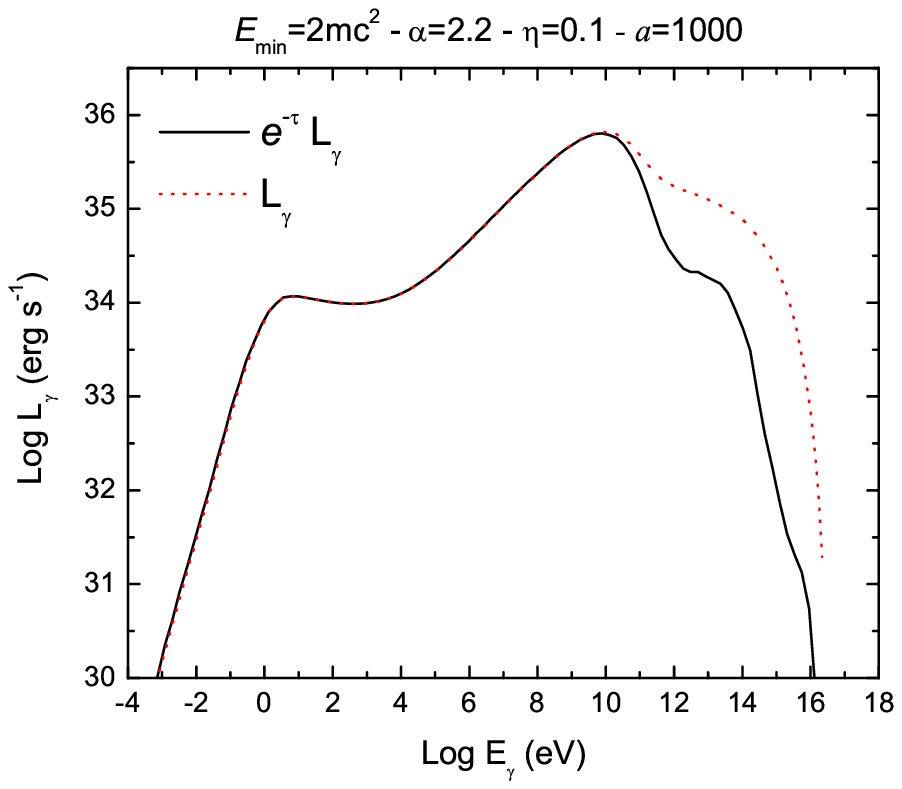}
\caption{Effects of the internal absorption in a `proton' microquasar (Romero \& Vila, in preparation).}
\label{fig4}
\end{figure}

The photo-hadronic model is based on interactions between relativistic protons in the jet and synchrotron emission from electrons and protons \cite{Levinson2001,Vila2008}. The high-energy photon density necessary for efficient gamma-ray production confines the acceleration and emission region to the base of the jet. Internal absorption is, then, very important, especially in cases where the photon field is dominated by electron synchrotron radiation \cite{Vila2008}. Instead, in models where the relativistic protons outnumber the electrons, high-energy gamma rays can escape with a spectrum basically unmodified by absorption. At high-energies, contributions from electron-positron pairs injected through photo-pair and photo-meson production can be important (see Fig. \ref{fig3}). Fig. \ref{fig4} shows a case with strong internal absorption.  

\section{The windy microquasar model}

In the case of HMMQs, external hadronic interactions can take place if there is some level of mixing between the jet flow and the stellar wind of the hot star. This is the original windy MQs model proposed by Romero et al. \cite{Romero2003}. Recent numerical simulations of jet-wind interactions show that the wind play an important role in the propagation and stability of the jet \cite{Ramon2008b}.   

\begin{figure}
\includegraphics[width=.8\textwidth]{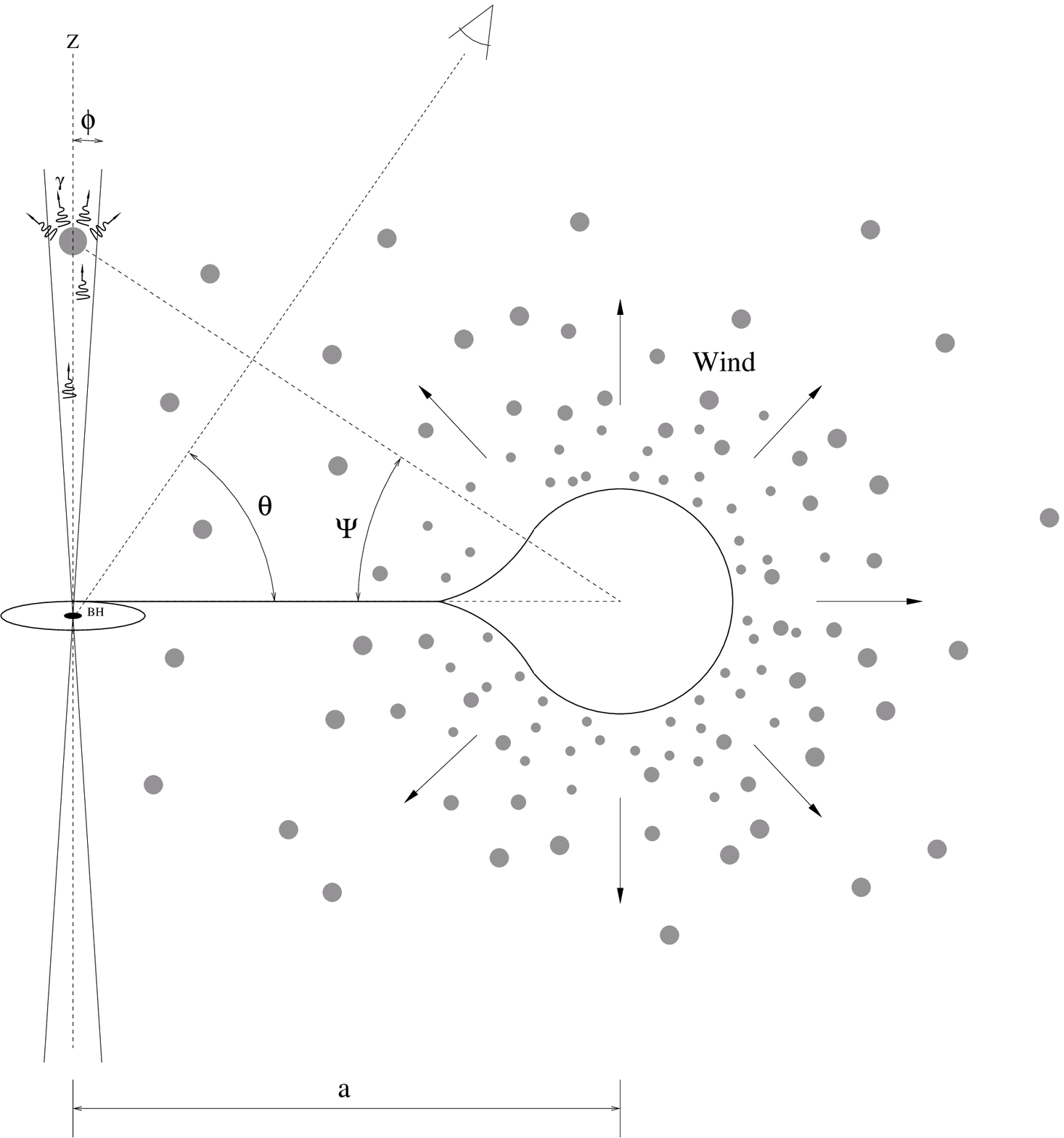}
\caption{Sketch of a HMMQ with a clumpy wind \cite{Romero2008}.}
\label{fig5}
\end{figure}

The wind of hot stars seems to have structure in the form of clumps, that are the result of plasma instabilities close to the stellar surface. These clumps have high densities (of the order of the density of the outer atmosphere) and propagate outwards embedded in the background wind. The density contrast can be as high as $10^{3-4}$. If some of these clumps interact with the jet, a rapid flare due to $pp$ collisions can arise \cite{Romero2008}. The situation is illustrated in Fig. \ref{fig5}.  

The interaction of a large number of clumps at different altitudes of the jet leads to a flickering of the gamma-ray light curve. If this flickering is measured by GLAST-Fermi satellite, its amplitude and associated time scales can be used to infer the properties of the clumping \cite{Owocki2008}. Typical values of the relative amplitude fluctuations would be $I_{\gamma}/\left\langle I_{\gamma} \right\rangle\sim 10$ \% \cite{Owocki2008}.

\section{Neutrinos}

Several authors have discussed neutrino production in microquasars (e.g. \cite{Romero2003,Levinson2001,Vila2008}). Romero \& Vila \cite{Vila2008}, who have taken into account internal photon absorption in heavy jets, notice that those sources with equipartition between relativistic electrons and protons (and hence strong synchrotron fields) are the most efficient for neutrino production above 1 TeV (the neutrinos are the result of photo-meson production). This means that not necessarily a strong neutrino source should be associated with a strong gamma-ray source. Another important point is that if the neutrinos are produced close to the base of the jet, where the magnetic field is very high, muon and pion losses can be important leading to significant neutrino attenuation \cite{Reynoso2009}. The most promising sources, then, are those where the hadronic interactions occur at some distance from the compact object as in \cite{Romero2003}. For additional discussion the reader is referred to Ref. \cite{Aha}.

\section{Polemical issues}

LS~I~+61~303 is a puzzling Be/X-ray binary with variable gamma-ray
emission up to TeV energies \cite{Albert2006}. The system was discovered by Gregory
\& Taylor \cite{Gregory1978}.  The primary star is of B0-B0.5Ve type. The orbital period is well-known from radio observations:
$P=26.4960$ days \cite{Gregory 2002}.  
The orbital parameters have been determined by Casares et al. \cite{Casares2005} and and Grundstrom et al. \cite{Grundstrom2007}.  
The orbit is inferred to be quite eccentric,
with eccentricity estimates from $e=0.72\pm0.15$ \cite{Casares2005} to
$e=0.55\pm0.05$ \cite{Grundstrom2007}. The inclination angle is unknown, and hence the nature of the compact object remains a mystery. There are different proposals to explain the non-thermal spectral energy distribution: a colliding wind scenario, that assumes the existence of a pulsar with a high spin-down luminosity \cite{Maraschi1981}, where particles are accelerated at the shock produced when the winds collide, and an accretion/ejection scenario where relativistic particles are accelerated close to the compact object \cite{Taylor1982}. For a detailed discussion see Ref. \cite{Romero2007}. Recently, Dhawan et al. \cite{Dhawan2006} presented a series of high-resolution radio-images of the source. In these images, during the periastron passage, a narrow feature is clearly seen pointing away from the position of the Be star. It was claimed that this can be interpreted as a `cometary' tail produced by the confinement of the pulsar wind by the wind of the Be star in accordance with the model presented in Ref. \cite{Dubus}. Extensive numerical simulations of (non-relativistic) colliding winds using a well-tested 3D SPH code \cite{Romero2007} and relativistic 2D simulations \cite{Bogovalov} showed, however, that it is extremely difficult to confine a pulsar wind with the weak wind from the star. In addition, if the wind of the star were powerful enough to produce the tail, Kelvin-Helmholtz and Rayleigh-Taylor instabilities would strongly affect the morphology during the periastron passage \cite{Romero2007}. If the spin-down luminosity of the pulsar is high enough to power the gamma-ray production, even with luminosities of $\sim 10^{36}$ erg s$^{-1}$ colliding wind models are on the verge of violating energy conservation in this system. A stronger pulsar wind would blow out the stellar wind in the opposite direction of what is observed. Amazingly, no consistent model able to explain both the radio-morphology and the global energetics has been developed by the advocates of the colliding wind scenario. Nonetheless, a significant part of the community seems to think that the case is closed and that LS~I~+61~303 is not an accreting source but a system similar to PSR B1259-63, a very different binary, with a much longer period (years) for which detailed models exist \cite{Tavani,Khangulyan2006}. 

The accretion simulations of LS~I~+61~303 show that the decretion disk of the Be star is truncated by tidal forces and the accretion rate is significantly sub-Eddington, but enough to power the whole emission. Moreover, the peak of the accretion rate is at phase $\sim 0.6$, where the maximum of the gamma-ray light curve is observed. The presence of strong density waves in the accretion disk makes the base of the jet unstable providing a simple explanation of the changing morphology. Romero et al. \cite{Romero2007} showed through extensive 3D SPH simulations that after 30 orbits a stationary state is reached. The power-law X-ray spectrum can be easily explained as synchrotron emission \cite{Ramon2006}, that erase any thermal feature from the spectrum.

Very recently, an extraordinary claim has been made: that the compact object in LS~I~+61~303 is a magnetar \cite{Dubus2}. It would be the only case of a magnetar in a binary system. This claim is at odds with the colliding wind scenario, since magnetars are very slow rotators with huge magnetic fields and are not expected to produce strong winds neither TeV gamma rays. It should be emphasized that the presence of a neutron star is \emph{not} a direct verification of the colliding wind model. Accretion onto pulsars with low magnetic fields can also power outflows as it is the case in Sco X-1 and similar sources (see \cite{Massi}). A number of potential counterparts for the recent flare observed in the direction of LS~I~+61~303 have been immediately found \cite{Paredes,Torres}.

It is undoubted that LS~I~+61~303 is a peculiar object that deserves more study. This study should be based on facts and detail modeling, avoiding prejudice and what the sociologists call `group thinking' \cite{Smolin}.

\section{Perspectives}

In the next couple of years GLAST-Fermi satellite, AGILE, and the new generation of Cherenkov telescope arrays (HESS II, MAGIC II, and VERITAS) will provide a broad sample of data that will allow to test the different types of MQ models and their predictions. A very exciting era is starting, where speculations will be replaced by detailed models, and these will survive only if they can confront the evidence. This is the way healthy science is.

\acknowledgments{I thank Prof. Emrah Kalemci for his kind invitation to attend this meeting and his support. I want to thank Felix Aharonian for his critical but constructive comments on my work along almost 10 years. Probably I have learned more physics from him than from any other scientist. I am also grateful to V. Bosch-Ramon, D. Khangulyan, Felix Mirabel, Mat\'{\i}as Reynoso, and Gaby Vila for fruitful discussions on microquasars, and to Paula Benagia for help with the manuscript. My work on high-energy astrophysics is supported by CONICET (PIP 5375) and the Argentine agency ANPCyT through Grant PICT 03-13291 BID 1728/OC-AC. Additional support comes from the Ministerio de Educaci\'on y Ciencia (Spain) under grant AYA 2007-68034-C03-01, FEDER funds.}

\end{document}